\documentclass[a4paper,11pt]{article}
\usepackage{pos}

\usepackage{algorithm}
\usepackage{algpseudocode}

\newcommand{\bx}{\mathbf{x}}
\newcommand{\by}{\mathbf{y}}

\newcommand{\bbC}{\mathbb{C}}
\newcommand{\bbR}{\mathbb{R}}

\newcommand{\calF}{\mathcal{F}}

\newcommand{\calO}{\mathcal{O}}

\newcommand{\calR}{\mathcal{R}}

\newcommand{\ReS}{\mathrm{Re}\,S}
\newcommand{\ImS}{\mathrm{Im}\,S}

\newcommand{\vev}[1]{\langle #1 \rangle}

\newcommand{\tr}{\mathrm{tr}\,}
\newcommand{\re}{\mathrm{Re}\,}

\makeatletter
\@addtoreset{equation}{section}

\def\@seccntformat#1{\csname the#1\endcsname.~~}
\makeatother

\title{
Analyzing the two-dimensional doped Hubbard model with the Worldvolume HMC method%
\footnote{Report No.: KUNS-3102}
}
\ShortTitle{Analyzing 2D doped Hubbard model with WV-HMC}

\author[a]{Masafumi Fukuma}
\author*[b]{Yusuke Namekawa}

\affiliation[a]{
Department of Physics, Kyoto University,
Kyoto 606-8502, Japan
}

\affiliation[b]{
Department of Computer Science, Fukuyama University,
Hiroshima 729-0292, Japan
}

\emailAdd{fukuma@gauge.scphys.kyoto-u.ac.jp}
\emailAdd{namekawa@fukuyama-u.ac.jp}

\abstract{
We apply the Worldvolume Hybrid Monte Carlo (WV-HMC) method [arXiv:2012.08468] to the two-dimensional Hubbard model, which is known to suffer from a severe sign problem when the system is doped (away from half filling). We show that the method predicts physical observables with controlled statistical errors on an $8 \times 8$ lattice at temperature $T/t = 1/6.4 \approx 0.156$ and interaction strength $U/t = 8.0$ ($t$ is the hopping amplitude), for which the standard determinant quantum Monte Carlo fails. 
}

\FullConference{The 42nd International Symposium on Lattice Field Theory (LATTICE2025)\\
2-8 November 2025\\
Tata Institute of Fundamental Research, Mumbai, India\\}

\begin{document}
%%%%%%%%%%%%%%%%%%%%%%%%%%%%%%%%%%%%%%%%%%%%%%%%%%%%%%%%%%%%%%%%%%%%%%%%
%%%%%%%%%%%%%%%%%%%%%%%%%%%%%%%%%%%%%%%%%%%%%%%%%%%%%%%%%%%%%%%%%%%%%%%%

\maketitle

%%%%%%%%%%%%%%%%%%%%%%%%%%%%%%
%%%%%%%%%%%%%%%%%%%%%%%%%%%%%%
\section{Introduction}
%%%%%%%%%%%%%%%%%%%%%%%%%%%%%%
%%%%%%%%%%%%%%%%%%%%%%%%%%%%%%

The numerical sign problem has long been a major obstacle 
to first-principle computations of various important physical systems, 
such as finite-density QCD, strongly correlated electron systems, 
and real-time dynamics of quantum many-body systems. 
Among them, the Hubbard model has a unique position in the research area 
of the sign problem, 
not only because of its importance in condensed-matter physics, 
but also because of its similarity 
in the mathematical structure to finite-density QCD. 

For the Hubbard model, 
the standard quantum Monte Carlo method 
suffers from the severe sign problem 
when the system is doped (away from half filling), 
and the model has been extensively studied using diverse methods,
ranging from Variational Monte Carlo~\cite{Yokoyama:1987,Yamaji:1998,Sorella:2005,Tahara:2008}
and Constrained-Path Auxiliary-Field Quantum Monte Carlo~\cite{Zhang:1995zz,Zhang:1996us},
to rather new approaches such as the Lefschetz thimble method~\cite{Mukherjee:2014hsa,Ulybyshev:2017hbs,Ulybyshev:2019hfm,Ulybyshev:2019fte,Ulybyshev:2022kxq,Ulybyshev:2024kdr},
the tempered Lefschetz thimble (TLT) method~\cite{Fukuma:2019wbv},
tensor renormalization group~\cite{Akiyama:2021xxr,Akiyama:2021glo},
complex-valued neural networks~\cite{Rodekamp:2022xpf},
constant path-integral contour shifts~\cite{Gantgen:2023byf},
and normalizing flows~\cite{Schuh:2026dvp}.

For the last fifteen years, 
there has been an academic trend 
to construct a versatile solution to the sign problem, 
and various methods have been proposed. 
Among them, the Worldvolume Hybrid Monte Carlo (WV-HMC) 
\cite{Fukuma:2020fez} 
(see also Refs.~\cite{Fukuma:2021aoo,Fukuma:2023eru,Fukuma:2025gya,
Fukuma:2025esu,Fukuma:2025uzg,Fukuma:2025cxg}) 
was proposed as a reliable and low-cost algorithm 
that resolves the sign problem. 
This is based on the idea of the thimble method 
but is free from the ergodicity problem inherent in thimble approaches 
by considering a continuum union of deformed integration surfaces. 
In this paper, based on Refs.~\cite{Fukuma:2025uzg,Fukuma:2025cxg}, 
we report its application to the two-dimensional doped Hubbard model. 
We particularly demonstrate that 
the method predicts physical observables with controlled statistical errors
on an $8 \times 8$ lattice 
at temperature $T/t = 1/6.4 \approx 0.156$ 
and interaction strength $U/t = 8.0$ 
($t$ is the hopping amplitude), 
for which the standard determinant quantum Monte Carlo fails.

%%%%%%%%%%%%%%%%%%%%%%%%%%%%%%
%%%%%%%%%%%%%%%%%%%%%%%%%%%%%%
\section{WV-HMC method via embedding GT-HMC}
%%%%%%%%%%%%%%%%%%%%%%%%%%%%%%
%%%%%%%%%%%%%%%%%%%%%%%%%%%%%%

We here give a brief review on the basics of WV-HMC. 
Let $x = (x^a) \in \bbR^N$ be a variable of $N$ degrees of freedom. 
Our aim is to numerically evaluate an observable $\calO(x)$ 
with a complex action $S(x)$: 
\begin{align}
  \langle \calO \rangle
  \equiv \frac{\int_{\bbR^N} dx \, e^{-S(x)} \, \calO(x)}
  {\int_{\bbR^N} dx \, e^{-S(x)}}
  \quad
  \Bigl(dx = dx^1 \wedge \cdots \wedge dx^N \equiv \prod_{a=1}^N dx^a \Bigr).
\label{vev}
\end{align}%
Following the Lefschetz thimble method 
\cite{Witten:2010cx,Cristoforetti:2012su,Cristoforetti:2013wha,
Fujii:2013sra,Fujii:2015bua,Fujii:2015vha,
Alexandru:2015xva,Alexandru:2015sua}, 
we complexify $x = (x^a)$ to $z = (z^i)$ ($i = 1,\ldots,N$), 
and deform the integration surface from $\Sigma_0 \equiv \bbR^N$ 
into $\Sigma$ within the complexified space $\bbC^N$. 
This deformation is governed by the anti-holomorphic flow: 
\begin{align}
  \dot{z} &= \overline{\partial S(z)}
  ~~\mbox{with}~~
  z|_{t=0} = x,
\label{flow_config}
\end{align}%
where $\dot{z} = \partial z/\partial t$ 
($t$: the deformation parameter referred to as the \emph{flow time}). 
Writing the solution by $z = z(t,x)$, 
the deformed surface at flow time $t$ is given by 
$\Sigma = \Sigma_t \equiv \{ z(t,x) \,|\, x \in \Sigma_0 \}$. 
Cauchy's theorem guarantees that the integrals do not change under the flow, 
so that Eq.~\eqref{vev} is written as 
\begin{align}
  \vev{\calO}
  = \frac{\int_{\Sigma_t} dz \, e^{-S(z)} \, \calO(z)}
  {\int_{\Sigma_t} dz \, e^{-S(z)}}
  \quad
  (dz = dz^1 \wedge \cdots \wedge dz^N).
\label{cauchy2}
\end{align}%
At large flow times $t$, 
$\Sigma_t$ approaches the Lefschetz thimbles 
(constant $\ImS$ surfaces), 
which suppresses the phase fluctuation from $e^{-i \ImS(z)}$.
However, this large $t$ introduces ergodicity issues: 
zeros of $e^{-S(z)}$ appear on $\Sigma_t$, 
acting as infinitely high potential barriers for Markov chain updates. 
The generalized thimble (GT) method \cite{Alexandru:2015sua} 
uses an intermediate $t$ 
to balance sign suppression and ergodicity. 
However, detailed studies show that 
sign suppression generally requires $\Sigma_t$ to crosses zeros 
\cite{Fukuma:2019wbv}, 
leaving the fundamental tension unresolved. 

The tempered Lefschetz thimble (TLT) method \cite{Fukuma:2017fjq} 
resolves this tension 
by treating $t$ as a dynamical variable. 
However, TLT incurs a high computational cost 
because it requires computing the deformation Jacobian 
at every replica exchange. 
The Worldvolume Hybrid Monte Carlo (WV-HMC) method 
\cite{Fukuma:2020fez} 
(see also Refs.~\cite{Fukuma:2021aoo,Fukuma:2023eru,Fukuma:2025gya,
Fukuma:2025esu,Fukuma:2025uzg,Fukuma:2025cxg}
eliminates this bottleneck.
This can be regarded as a continuous version of the TLT method 
and introduced as follows. 

Since both the numerator and the denominator 
in Eq.~\eqref{cauchy2} are independent of $t$ 
(due to Cauchy's theorem),
we can take averages over $t$ separately 
with an arbitrary common weight $e^{-W(t)}$ \cite{Fukuma:2020fez}:
\begin{align}
  \vev{\calO}
  = \frac{\int dt \, e^{-W(t)} \int_{\Sigma_t} dz \, e^{-S(z)}\,\calO(z)}
  {\int dt \, e^{-W(t)} \int_{\Sigma_t} dz \, e^{-S(z)}} .
\label{observable_WV_HMC}
\end{align}%
This takes the form of a ratio of 
reweighted averages on the \emph{worldvolume} $\calR$,
\begin{align}
  \calR\equiv \bigcup_{t} \Sigma_t = \{z(t,x) \mid t\in\bbR,\,x\in\bbR^N\}, 
\end{align}%
as 
\begin{align}
  \vev{\calO} 
  &= \frac{\vev{ \calF_\calR(z)\,\calO(z) }_\calR}
  {\vev{ \calF_\calR(z) }_\calR},
\label{rewt_calR}
\\
  \vev{g(z)}_\calR
  &\equiv \frac{\int_\calR |dz|_\calR \, e^{-\ReS(z) - W(t)}\,g(z)}
  {\int_\calR |dz|_\calR \,e^{-\ReS(z) - W(t)}}.
\label{vev_calR1}
\end{align}%
Here, $|dzr|_\calR$ is the invariant measure on $\calR$, 
and $\calF_\calR(z) \equiv dt\,dz\,e^{-i\,\ImS(z)} / |dz|_\calR$ 
is the associated reweighting factor. 
The extent of the worldvolume $\calR$ in the $t$-direction 
can be restricted within a finite interval $[T_0,T_1]$ 
by tuning the function $W(t)$, 
which we take as follows \cite{Fukuma:2023eru}:
\begin{align}
  W(t) = 
  \begin{cases}  
    -\,\gamma(t-T_0) + c_0\,\bigl(e^{(t-T_0)^2/2d_0^2} - 1\bigr) 
    &  \mbox{\ for \ } t < T_0              
    \\
    -\,\gamma(t-T_0)                                             
    &  \mbox{\ for \ } T_0 \leq t \leq T_1  
    \\
    -\,\gamma(t-T_0) + c_1\,\bigl(e^{(t-T_1)^2/2d_1^2} - 1\bigr) 
    &  \mbox{\ for \ } t > T_1 .            
    \\
  \end{cases}
\label{Wt}
\end{align}
The reweighted average $\vev{g(z)}_\calR$ can be further rewritten 
as a phase-space integral over the tangent bundle of $\calR$, 
\begin{align}
  T\calR = \{ (z,\pi) \,|\, z \in \calR,\, \pi \in T_z \calR \},
\end{align}
as 
\begin{align}
  \vev{g(z)}
  = \frac{
       \int_{T\calR} d\Omega_\calR\,e^{-H(z,\pi)}\,g(z)
    }{
       \int_{T\calR} d\Omega_\calR\,e^{-H(z,\pi)}
    }.
\end{align}
Here, $d\Omega_\calR \equiv \omega_\calR^{N+1} / (N+1)!$ 
is the symplectic volume form constructed 
from the symplectic 2-form $\omega_\calR = \re ( d\pi^\dagger\wedge dz)$, 
and $H(z,\pi) = (1/2)\,\pi^\dagger \pi + \ReS(z) + W(t(z))$.%
\footnote{ %-----
  $t(z)$ returns the flow time $t$ for configuration $z = z(t,x)$.
} %--------------
The distribution function $\propto e^{-H(z,\pi)}$ 
can be generated using the HMC algorithm 
with constrained molecular dynamics such as RATTLE \cite{Andersen:1983}. 
See Refs.~\cite{Fukuma:2020fez,Fukuma:2023eru,Fukuma:2025gya} 
for details. 

An HMC algorithm for the GT method 
[which we refer to as the \emph{generalized thimble HMC} (GT-HMC)]
can be introduced similarly 
\cite{Alexandru:2019,Fukuma:2019uot} 
(see also Ref.~\cite{Fukuma:2023eru})
as a sampling on the tangent bundle of $\Sigma$,
$T\Sigma = \{ (z,\pi) \,|\, z\in\Sigma,\,\pi\in T_z \Sigma \}$,  
from the distribution $\propto e^{-H(z,\pi)}$ 
with the Hamiltonian $H(z,\pi) = (1/2)\,\pi^\dagger \pi + \ReS(z)$. 
While GT-HMC alone suffers from ergodicity issues, 
embedding it as a subprocess within WV-HMC 
strongly enhances global sampling, 
particularly when the worldvolume is a thin layer \cite{Fukuma:2025cxg}.

%%%%%%%%%%%%%%%%%%%%%%%%%%%%%%
%%%%%%%%%%%%%%%%%%%%%%%%%%%%%%
\section{The Hubbard model}
%%%%%%%%%%%%%%%%%%%%%%%%%%%%%%
%%%%%%%%%%%%%%%%%%%%%%%%%%%%%%

The Hubbard model on a $d$-dimensional spatial lattice 
is defined by the following Hamiltonian 
(including the chemical potential term): 
\begin{align}
  \hat H_\mu^{\text{org}}
  &= \hat H - \mu \hat N
\nonumber
\\
  &\equiv
  -\sum_{\bx, \by} t_{\bx \by} \sum_{\sigma=\uparrow,\downarrow} 
    c_{\bx, \sigma}^\dag c_{\by, \sigma}
  + U      \sum_{\bx} n_{\bx, \uparrow} n_{\bx, \downarrow} 
  - \mu    \sum_{\bx} (n_{\bx, \uparrow} + n_{\bx, \downarrow}) .
\label{hamiltonian_org}
\end{align}%
Here,  $c_{\bx, \sigma}$ and $c_{\bx, \sigma}^\dag$ denote  
the annihilation and creation operators, respectively, 
of an electron with spin $\sigma\,(=\uparrow,\,\downarrow)$ 
at site $\bx=(x_i)$ $(i=1,\ldots,d)$, 
and $n_{\bx, \sigma} \equiv c_{\bx, \sigma}^\dag c_{\bx, \sigma}$.  
$t = (t_{\bx \by})$ is the hopping matrix, 
where $t_{\bx\by} = t\,(>0)$ if $\bx$ and $\by$ are nearest neighbors,
and $t_{\bx\by} = 0$ otherwise.
$U$ is the on-site repulsion strength, 
and $\mu$ is the chemical potential associated with the number operator 
$\hat N = \sum_\bx \sum_\sigma n_{\bx,\sigma}$.
We assume that the model is defined on a periodic, bipartite square lattice 
of linear size $L_s$, 
so that the spatial volume is given by $V_d \equiv L_s^d$. 

We perform a particle-hole transformation on the down-spin component 
and write 
\begin{align}
  a_\bx \equiv c_{\bx\uparrow},\quad
  b_\bx \equiv (-1)^\bx c^\dag_{\bx\downarrow},
\end{align}%
where $(-1)^\bx\equiv (-1)^{\sum_i x_i}$ denotes the parity of site $\bx$.
Under this transformation, 
the Hamiltonian \eqref{hamiltonian_org} becomes 
(up to an additive constant $-\mu V_d$):
\begin{align}
  \hat H_\mu 
  \equiv \hat{H}_\mu^{\text{org}} - \mu V_d
  = -\sum_{\bx, \by} 
  t_{\bx \by}\, ( a_\bx^\dag a_\by + b_\bx^\dag b_\by )
  + \frac{U}{2}\, \sum_{\bx} ( n^a_\bx - n^b_\bx )^2
  - \tilde \mu\, \sum_{\bx} ( n^a_\bx - n^b_\bx ) ,
\end{align}%
where $n^a_\bx \equiv a_\bx^\dag a_\bx$ 
and $n^b_\bx \equiv b_\bx^\dag b_\bx$, 
and the shifted chemical potential is defined as 
\begin{align}
  \tilde \mu \equiv \mu - \frac{U}{2}.
\end{align}%
The point $\mu=U/2$ (i.e., $\tilde\mu=0$) corresponds to half filling, 
where $\vev{ n_{\bx,\uparrow} + n_{\bx,\downarrow} } = 1$ 
(i.e., $\vev{ n^a_\bx - n^b_\bx } = 0$). 

Following Ref.~\cite{Beyl:2017kwp}, 
we introduce the redundant parameter $\alpha$ $(0\leq\alpha\leq 1)$ as% 
\footnote{ %-----
  The equality directly follows from the identity 
  $ (n^a_\bx + n^b_\bx - 1)^2 = -(n^a_\bx - n^b_\bx)^2 + 1$ 
  (note that $(n^{a/b}_\bx)^2 = n^{a/b}_\bx$)
  \cite{Beyl:2017kwp}. 
} %--------------
\begin{align}
  (n^a_\bx - n^b_\bx)^2 = \alpha (n^a_\bx - n^b_\bx)^2
  - (1-\alpha) (n^a_\bx + n^b_\bx -1)^2 + 1 - \alpha.
\label{alpha}
\end{align}%
We then complete the square 
by using two auxiliary variables (Hubbard-Stratonovich variables): 
\begin{align}
  &e^{-(\alpha\epsilon U/2)\,(n^a - n^b)^2 
      + ((1-\alpha)\epsilon U/2)\,(n^a + n^b - 1)^2 
      - (1-\alpha)\epsilon U/2}
\nonumber
\\
  &= \int dA dB\, e^{-(1/2)(A^2 + B^2)}\,
  e^{ [ i c_0 A + c_1 B - c_1^2] \, n^a} \,
  e^{ [-i c_0 A + c_1 B - c_1^2] \, n^b} 
\end{align}%
with $c_0 \equiv \sqrt{\alpha\epsilon U}$ 
and $c_1 \equiv \sqrt{(1-\alpha)\epsilon U}$. 
We decompose the inverse temperature $\beta$ into $N_t$ time slices 
and introduce a spacetime lattice of volume 
$V_{d+1} \equiv N_t V_d = N_t\times L_s^d$, 
whose coordinates are labeled by $x = (\ell,\bx)$ 
$(\ell = 1,\ldots,N_t)$.   
Then, the grand canonical partition function is given as follows 
(see Ref.~\cite{Fukuma:2025uzg} for derivation):  
\begin{align}
  Z = \int dA\,dB\,e^{-S(A,B)}
  = \int dA\,dB\,e^{-S_0(A,B)}\,\det D_a(A,B)\,\det D_b(A,B).
\label{path_int_AB}
\end{align}% 
Here, $A = (A_x)$ and $B = (B_x)$ are scalar fields on the spacetime lattice, 
and we have introduced 
$V_{d+1} \times V_{d+1}$ matrices 
$t = (t_{xy})$ and $\Lambda_0 = ((\Lambda_0)_{xy})$ 
with indices $x = (\ell, \bx)$ and $y= (m, \by)$
(we reuse the symbol $t$),
\begin{align}
  t_{xy} \equiv \delta_{\ell m}\,t_{\bx\by},
  \quad
  (\Lambda_0)_{xy} \equiv 
   \left\{ \begin{array}{ll}
      \delta_{\ell+1,m}\,\delta_{\bx\by} & (\ell < N_t) \\
      - \delta_{1,m}\,\delta_{\bx\by}      & (\ell = N_t).
   \end{array}\right. 
\end{align}% 
Moreover, $S_0(A,B) \equiv (1/2)\sum_x (A_x^2 + B_x^2)$, 
$h_{a/b} = ((h_{a/b})_x)$ are diagonal matrices with 
\begin{align}
  (h_{a/b})_x = e^{\pm (\epsilon\tilde\mu + i c_0 A_x) + c_1 B_x - c_1^2}, 
\end{align}%
and $D_{a/b}$ are fermion matrices, 
\begin{align}
  D_{a/b}(A,B) \equiv 
  h_{a/b} - e^{-\epsilon t}\,\Lambda_0.
\end{align}%
We employ the symmetric Trotter decomposition, 
which matches the continuum evolution operator $e^{-\epsilon \hat{H}_\mu}$ 
up to $O(\epsilon^2)$. 
Accordingly, 
we expand $D_{a/b}$ to the same order:%
\footnote{ %-----
  Note that 
  $\Lambda_0 = 1 + O(\epsilon)$ holds only for thermalized configurations 
  and should not be used as a general estimate. 
} %-------------- 
\begin{align}
  D_{a/b} = h_{a/b} - \Lambda_0 + \epsilon t \Lambda_0
    - \frac{\epsilon^2}{2}\,t^2 \Lambda_0.
\end{align}%
As discussed in Ref.~\cite{Fukuma:2025uzg}, 
the identity $D_b = D_a^\ast$ holds at half filling ($\tilde\mu = 0$), 
yielding $\det D_a\, \det D_b = |\det D_a|^2$.
This ensures that 
the path integral is free from the sign problem at half filling.
We expect the sign problem to remain mild when $D_b \approx D_a^\ast$, 
which occurs for small $\alpha$. 
However, choosing $\alpha$ too small introduces ergodicity issues: 
zeros of $\det D_{a/b}$ appear 
on or near the original configuration space $\Sigma_0$, 
as detailed in Ref.~\cite{Beyl:2017kwp}.
Thus, there exists an optimal $\alpha$ 
that mitigates the sign problem on $\Sigma_0$ 
without triggering ergodicity issues.

We define 
the number density operator $n$ 
and the energy density operator $e$ 
as follows \cite{Fukuma:2025uzg}:
\begin{align}
  n(A,B) 
  &\equiv 
  - \frac{1}{V_{d+1}}\,
    \frac{\partial S(A,B)}{\partial (\epsilon\mu)}\Bigr|_{\epsilon}
  + 1
  = - \frac{1}{V_{d+1}}\,
    \frac{\partial S(A,B)}{\partial (\epsilon\tilde\mu)}\Bigr|_{\epsilon}
  + 1 ,
\label{number_density}
\\
  e(A,B) 
  &\equiv 
  \frac{\partial S(A,B)}{\partial \epsilon}\Bigr|_{\epsilon\mu}
  = \frac{1}{V_{d+1}}\,\biggl[
  \frac{\partial S(A,B)}{\partial \epsilon}\Bigr|_{\epsilon\tilde\mu}
  - \frac{U}{2}\,
  \frac{\partial S(A,B)}{\partial (\epsilon\tilde\mu)}\Bigr|_{\epsilon}
  \biggr] .
\label{energy_density}
\end{align}%
Their expectation values can be estimated via the path integral, 
and are expected to agree with the continuum expectation values 
of $\hat N/V_d$ and $\hat H/V_d$ 
up to $O(\epsilon^2)$ corrections: 
\begin{align}
  \langle n \rangle &\equiv 
  \frac{1}{V_{d+1}}\,\frac{\int (dA\,dB)\,e^{-S(A,B)}\,n(A,B)}
       {\int (dA\,dB) e^{-S(A,B)}}
  = \frac{1}{V_d}\,\frac{\tr\, e^{-\beta (\hat H - \mu \hat N)}\,\hat N }
                        {\tr\, e^{-\beta (\hat H - \mu \hat N)} }
    + O(\epsilon^2) ,
\label{scaling_n}
\\
 \langle e \rangle 
 &\equiv 
 \frac{1}{V_{d+1}}\,\frac{\int (dA\,dB)\,e^{-S(A,B)}\,e(A,B)}
 {\int (dA\,dB) e^{-S(A,B)}}
 = \frac{1}{V_d}\,\frac{\tr\, e^{-\beta (\hat H - \mu \hat N)}\,\hat H }
 {\tr\, e^{-\beta (\hat H - \mu \hat N)} }
 +  O(\epsilon^2). 
\label{scaling_e}
\end{align}%

%%%%%%%%%%%%%%%%%%%%%%%%%%%%%%
%%%%%%%%%%%%%%%%%%%%%%%%%%%%%%
\section{Setup}
%%%%%%%%%%%%%%%%%%%%%%%%%%%%%%
%%%%%%%%%%%%%%%%%%%%%%%%%%%%%%

We apply WV-HMC to the two-dimensional doped Hubbard model 
on a lattice of size $L_s \times L_s = 8 \times 8$ 
with the following parameters:  
hopping amplitude $t = 1.0$, 
inverse temperature $\beta = 6.4$, 
repulsion strength $U = 8.0$ 
for various values of the chemical potential $\mu$.%
\footnote{ %-----
  In the common notation in condensed matter physics, 
  these parameters correspond to 
  $T/t = 1/(t\beta) = 1/6.4 \simeq 0.156$ and $U/t = 8.0$. 
} %--------------
The Trotter numbers are $N_t=24, 22, 20, 18$,
corresponding to the Trotter steps $\epsilon = 0.27,\,0.29,\,0.32,\,0.36$. 
We tune $\alpha$ to the smallest value 
that avoids ergodicity issues. 
This allows us to keep the maximum flow time $T_1$ as small as possible. 
We choose the upper cutoff $T_1$ 
such that the average phase factor on $\Sigma_t$, 
computed using GT-HMC, 
becomes statistically distinguishable from zero 
at the two-sigma level 
at $t \sim T_1$.
The weight function parameters [see Eq.~\eqref{Wt}] 
are set as follows: 
$\gamma = 0$, $c_0 = c_1 = 0.01$, and $d_0 = d_1 = 2.0\times 10^{-3}$.

We perform the simulations using a combined update: 
two sets of embedded GT-HMC 
(each trajectory consisting of 25 MD steps  
with $\Delta s = 4.0 \times 10^{-2}$) 
followed by one set of pure WV-HMC 
(each trajectory consisting of 25 MD steps 
with $\Delta s = 4.0 \times 10^{-4}$). 
Observables are measured after each combined update, 
and statistical errors are estimated 
using the blocked jackknife method.

%%%%%%%%%%%%%%%%%%%%%%%%%%%%%%
%%%%%%%%%%%%%%%%%%%%%%%%%%%%%%
\section{Results}
%%%%%%%%%%%%%%%%%%%%%%%%%%%%%%
%%%%%%%%%%%%%%%%%%%%%%%%%%%%%%

\begin{figure}[ht]
  \centering
  \includegraphics[width=49mm]{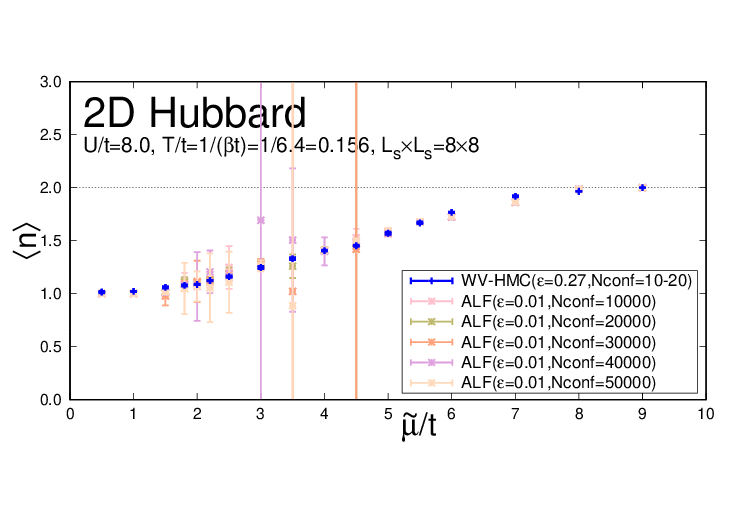}
  \includegraphics[width=49mm]{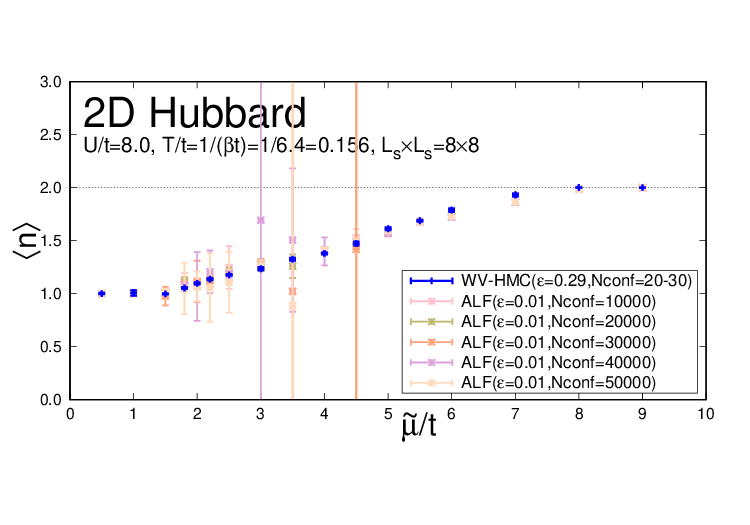}
  \includegraphics[width=49mm]{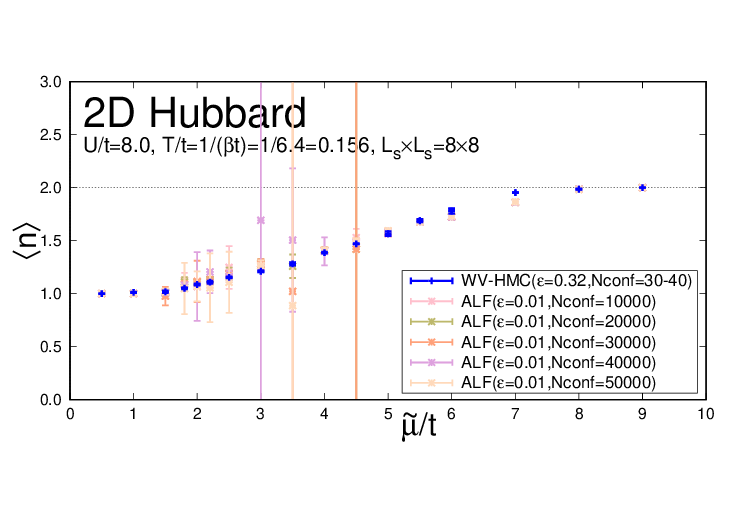}
  \caption{%
    The number density $\vev{n}$
    at $\epsilon= 0.27, 0.29, 0.32$,
    compared with results using ALF at $\epsilon= 0.01$.
  }
\label{fig:observables_8x8}
\end{figure}

Figure~\ref{fig:observables_8x8}
shows the number density as a function of $\tilde{\mu}$ 
at the Trotter steps $\epsilon = 0.27,\,0.29,\,0.32$
with $\beta$ fixed to 6.4.
Results obtained using ALF (Algorithm for Lattice Fermions) 
\cite{Bercx:2017pit,ALF:2020tyi} 
at $\epsilon = 0.01$ are also shown for comparison.
We see that 
WV-HMC evaluates the number density with controlled statistical errors
in a parameter region where the sign problem is serious.
We also observe discrepancies between WV-HMC and ALF results
in the region $\tilde{\mu} = 5.5 \text{--} 7.0$, 
for which the sign problem is not serious 
and thus the ALF results are reliable.

\begin{figure}[ht]
  \centering
  \includegraphics[width=36mm]{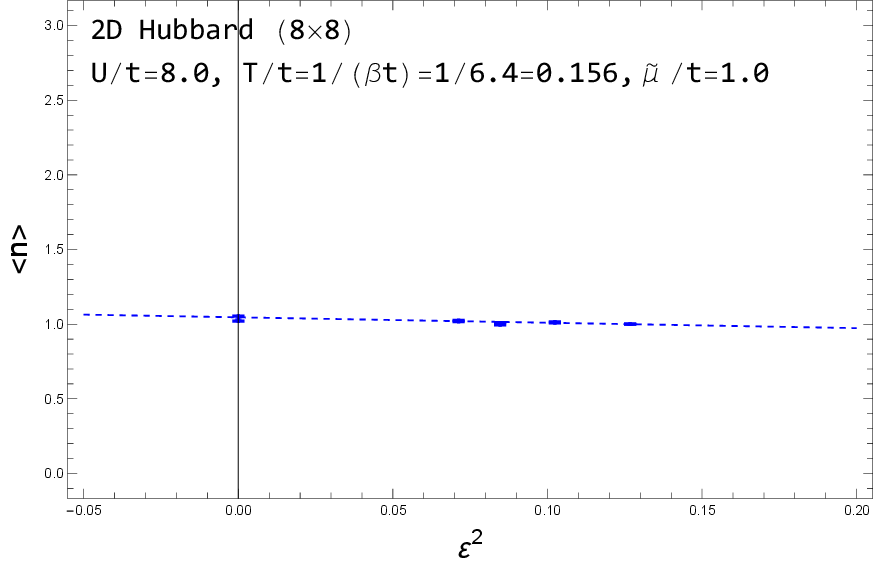}
  \includegraphics[width=36mm]{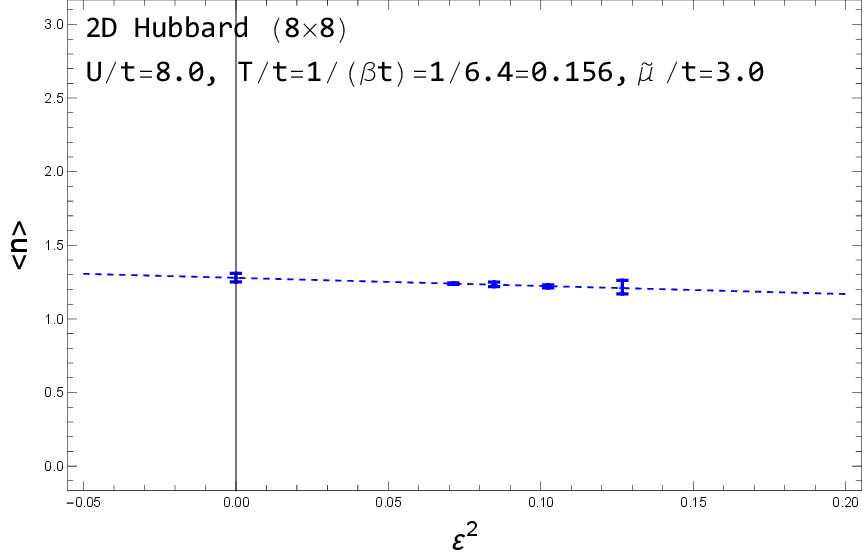}
  \includegraphics[width=36mm]{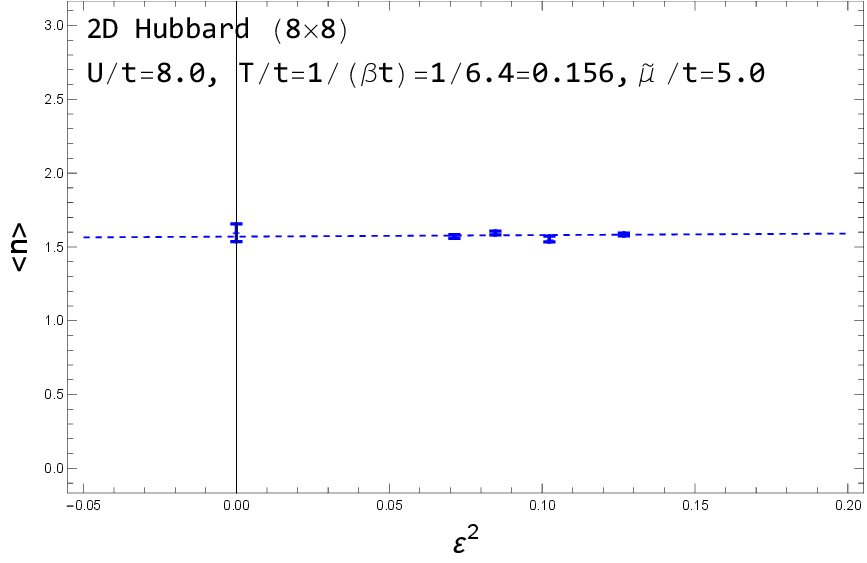}
  \includegraphics[width=36mm]{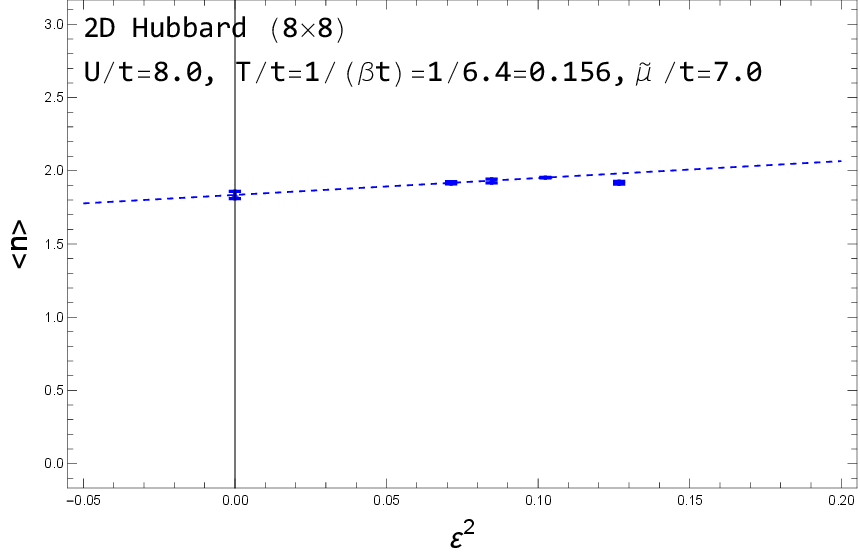}
  \caption{%
    Finite-$\epsilon$ effects 
    in the number density $\vev{n}$
    for $\tilde\mu = 1.0,\,3.0,\,5.0,\,7.0$
    (figure adapted from \cite{Fukuma:2025cxg}).
  }
\label{fig:observables_8x8_eps_scaling}
\end{figure}

\begin{figure}[ht]
  \centering
  \includegraphics[width=70mm]
    {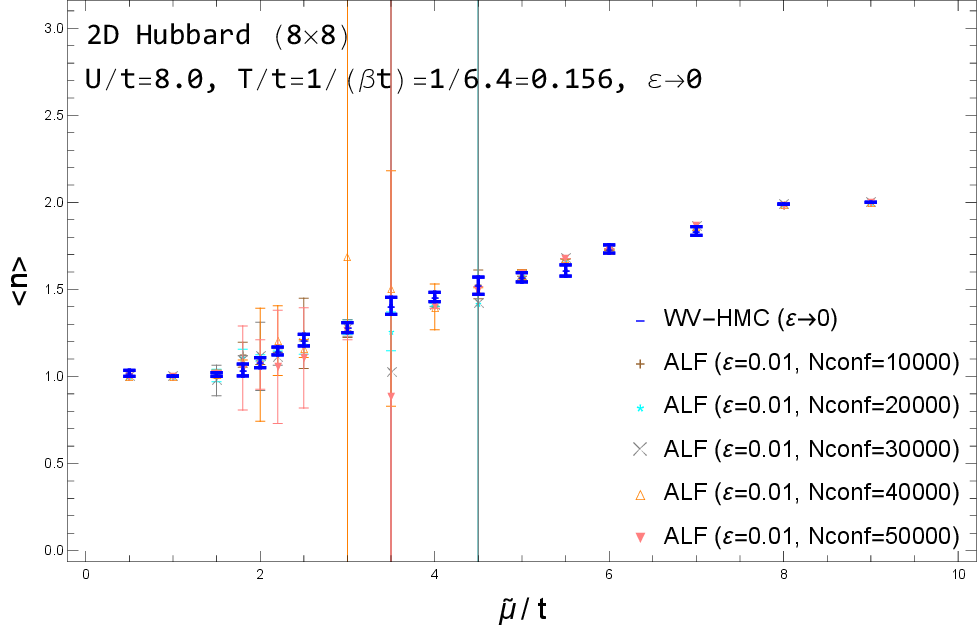}
  %\vspace{-10ex}\\
  \hspace{3ex}
  \includegraphics[width=70mm]
    {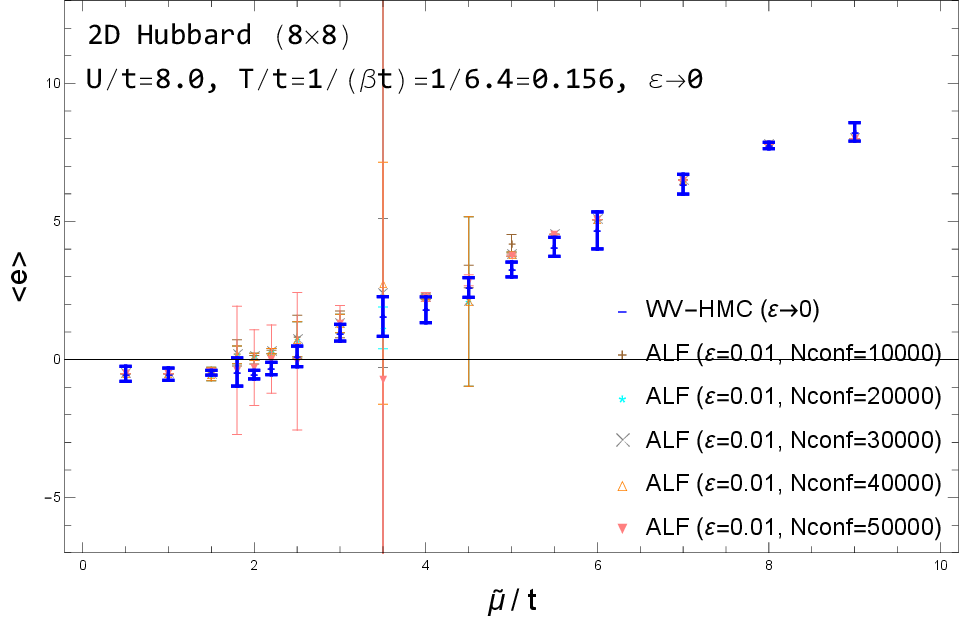}
  \caption{%
    Number density $\vev{n}$ and energy density $\vev{e}$ 
    in the continuum limit $\epsilon \to 0$,
    compared with results using ALF at $\epsilon= 0.01$
    (figure adapted from \cite{Fukuma:2025cxg}).
  }
\label{fig:observables_8x8_wv_vs_alf}
\end{figure}

Figure~\ref{fig:observables_8x8_eps_scaling} shows that 
the number density $\vev{n}$ can be fit as $a + b \epsilon^2$
for the results at $\epsilon = 0.27,\,0.29,\,0.32$ 
(as well as the result at $\epsilon = 0.36$ 
when it is consistent with the linear trend),
which agrees with the expectation 
from Eqs.~\eqref{scaling_n} and \eqref{scaling_e}. 
The energy density $\vev{e}$ can also be fit with the same functional form. 
These fits yield the observables 
in the continuum limit $\epsilon \to 0$,
which are shown in Fig.~\ref{fig:observables_8x8_wv_vs_alf}. 
We find that WV-HMC yields observables with controlled statistical errors 
even in the continuum limit.
The discrepancies between WV-HMC and ALF results observed at finite $\epsilon$ 
indeed disappear in the continuum limit.

%%%%%%%%%%%%%%%%%%%%%%%%%%%%%%
%%%%%%%%%%%%%%%%%%%%%%%%%%%%%%
\section{Conclusion}
%%%%%%%%%%%%%%%%%%%%%%%%%%%%%%
%%%%%%%%%%%%%%%%%%%%%%%%%%%%%%

We applied WV-HMC to the two-dimensional doped Hubbard model 
on an $8 \times 8$ lattice at $T/t = 1/6.4 \simeq 0.156$ and $U/t = 8.0$. 
The tuning of $\alpha$ mitigates the sign problem 
on the original integration surface, 
allowing the maximum flow time $T_1$ to be significantly reduced. 
Although this introduces an ergodicity issue for a thin layer of WV-HMC, 
the problem is successfully resolved by embedding GT-HMC into the process. 
Using this combined algorithm, 
we took the continuum limit in the temporal direction ($\epsilon \to 0$) 
and demonstrated that observables can be evaluated 
with controlled statistical errors.
The techniques developed here are expected to scale 
for the Hubbard model on larger lattices, 
which we are currently investigating.

%%%%%%%%%%%%%%%%%%%%%%%%%%%%%%
%%%%%%%%%%%%%%%%%%%%%%%%%%%%%%
\acknowledgments
The authors thank Sinya Aoki, Fakher F.\ Assaad, Masatoshi Imada, Ken-Ichi Ishikawa, Issaku Kanamori, Yoshio Kikukawa, Nobuyuki Matsumoto, Yusuke Nomura, Maksim Ulybyshev, Youhei Yamaji, and Shiwei Zhang for valuable discussions.
This work was partially supported by JSPS KAKENHI Grant Numbers 
JP20H01900, JP21K03553, JP23H00112, JP23H04506, JP24K07052; 
by MEXT as "Program for Promoting Researches on the Supercomputer Fugaku" 
(Simulation for basic science: approaching the new quantum era, JPMXP1020230411); 
by SPIRIT2 2025 of Kyoto University. 
We used computational resources of the supercomputer Fugaku 
provided by the RIKEN Center for Computational Science 
(Project ID: hp230207, hp240213).
%%%%%%%%%%%%%%%%%%%%%%%%%%%%%%
%%%%%%%%%%%%%%%%%%%%%%%%%%%%%%

%%%%%%%%%%%%%%%%%%%%%%%%%%%%%%
%%%%%%%%%%%%%%%%%%%%%%%%%%%%%%
\bibliographystyle{JHEP}
\bibliography{ref.bib}
%%%%%%%%%%%%%%%%%%%%%%%%%%%%%%
%%%%%%%%%%%%%%%%%%%%%%%%%%%%%%

\end{document}